\documentclass[5p,twocolumn,times,number]{elsarticle}

\usepackage{graphicx}
\usepackage{amsmath}   
\def\pmbanner{{\hrule height 1 pt}\vskip35pt{NIMA POST-PROCESS BANNER TO BE REMOVED AFTER FINAL ACCEPTANCE}\vskip35pt{\hrule height 4pt}\vskip20pt}
\begin{document}

\begin{frontmatter}

%% Note: \pmbanner before the actual title
\title{\pmbanner Detector setup of the VIP2 Underground Experiment at LNGS}

\author[add1]{J.~Marton\corref{cor}}
\author[add1]{A.~Pichler}
\ead{johann.marton@oeaw.ac.at}
\author[add12]{H. Shi}
\author[add3]{E. Milotti}
\author[add2]{S. Bartalucci}
\author[add2]{M. Bazzi}
\author[add4]{S. Bertolucci}
\author[add2]{A.M. Bragadireanu}
\author[add1]{M. Cargnelli}
\author[add2]{A. Clozza}
\author[add2]{C. Curceanu}
\author[add2]{L. De Paolis}
\author[add6]{S. Di Matteo}
\author[add7]{J.-P. Egger}
\author[add8]{H. Elnaggar}
\author[add2]{C. Guaraldo}
\author[add2]{M. Iliescu}
\author[add9]{M. Laubenstein}
\author[add2]{M. Miliucci}
\author[add2]{D. Pietreanu}
\author[add2]{K. Piscicchia}
\author[add2]{A. Scordo}
\author[add2]{D.L. Sirghi}
\author[add2]{F. Sirghi}
\author[add2]{L. Sperandio}
\author[add2]{O. Vazquez Doce}
\author[add1]{E. Widmann}
\author[add1]{J. Zmeskal}

%\author[add3]{on behalf of the VIP2 Collaboration}
%\author[add6]{M.~Author3}

\cortext[cor]{Johann Marton}

\address[add1]{Stefan-Meyer-Institut f\"ur Subatomare Physik, Boltzmanngasse 3, 1090 Wien, Austria}
\address[add2]{INFN, Laboratori Nazionali di Frascati, C.P. 13, Via E. Fermi 40, I-00044 Frascati(Roma), Italy}
\address[add3]{Dipartimento di Fisica, Universita di Trieste and INFN-Sezione di Trieste, Via Valerio, 2, I-34127 Trieste, Italy}
\address[add4]{Dipartimento di Fisica e Astronomia, Universita di Bologna, Viale Berti Pichat 6/2, Bologna, Italy}
%\address[add5]{IFIN-HH, Institutul National pentru Fizica si Inginerie Nucleara Horia Hulubbei, Reactorului 30, Magurele, Romania}
\address[add6]{Universite de Rennes, CNRS, IPR (Institute de Physique de Rennes), UMR 6251, F-35000 Rennes, France}
\address[add7]{Institut de Physique, Universite de Neuchatel, 1 rue A.-L. Breguet, CH-2000 Neuchatel, Switzerland}
\address[add8]{Debye Institute for Nanomaterial Science - Utrecht University, P.O. Box 80.000 3508 TA Utrecht, The Netherlands}
\address[add9]{INFN, Laboratori Nazionali del Gran Sasso, I-67010 Assergi (AQ), Italy}
%\address[add10]{Museo Storico della Fisica e Centro Studi e Ricerche Enrico Fermi, Piazza del Viminale 1, 00183 Roma, Italy}
%\address[add11]{Excellence Cluster Universe, Technische Universit ̈at M ̈unchen, Boltzmannstraße 2, D-85748 Garching, Germany}
\address[add12]{Institut f\"ur Hochenergiephysik der \"Osterreichischen Akademie der Wissenschaften, Nikolsdorfer Gasse 18, 1050 Wien, Austria}

\begin{abstract}
The VIP2 experiment tests the Pauli Exclusion Principle with high sensitivity, by searching for Pauli-forbidden atomic transitions from the 2p to the 1s shell in copper at about 8keV. The transition energy of Pauli-forbidden Kα X-rays is shifted by about 300 eV with respect to the normal allowed Kα line. This energy difference can be resolved using Silicon Drift Detectors. The data for this experiment is taken in the Gran Sasso underground laboratory (LNGS), which provides shielding from cosmic radiation.
An overview of the detection system of the VIP2 experiment will be given. This includes the Silicon Drift Detectors used as X-ray detectors which provide an energy resolution of around 150 eV at 6 keV and timing information for active shielding. Furthermore, the low maintenance requirement makes them excellent X-ray detectors for the use in an underground laboratory. The VIP2 setup will be discussed which consists of a high current target system and a passive as well as an active shielding system using plastic scintillators read out by Silicon Photomultipliers.

\end{abstract}


\end{frontmatter}

\section{Introduction}
Wolfgang Pauli formulated the Pauli Exclusion Principle (PEP) in 1925
explaining the shell structure of atoms. It turned out that this principle is connected to the spin-statistics theotrem and valid not
only for electrons - it is valid for all fermions, i. e. particles with half integer spin.
In spite of the overwhelming success of the PEP in explaining many features of
nature, a loophole-free proof cannot be given up to now.
The experiment VIP2\footnote{International Collaboration "VIolation of the Pauli Principle" at LNGS, Gran Sasso} employ a method to test the PEP similar to that of Ramberg
and Snow \cite{Ramberg1990}. By circulating an electric current fresh electrons are inserted into
a Cu strip. They form new quantum states with atoms of the conductor. These
quantum states have a probability of to be non-Paulian. The electron then
cascades to the 1s state and thereby emits photons from non-Paulian transitions,
which are shifted in energy by about 300 eV. This shifted X-rays can be resolved by spectroscopy with Silicon Drift Detectors (SDDs) used as X-ray detectors. The
number of possible photons from these transitions, which are identified by their
energy, is used to set an upper limit for the probability for a violation of the PEP. A small violation of the PEP is qualitatively described by the quantity $\beta^{2}$/2, which
can be traced back to a model introduced by Ignatiev and Kuzmin \cite{Ignatiev87} and is commonly used in the literature.

\section{VIP2 Setup and Detectors}

Silicon Drift Detectors (SDDs) are semiconductor detectors ideally suited for soft X-ray spectroscopy. Free
electrons generated by incident radiation drift to the anode due to an applied electric field. From the
number of electrons at the anode, the energy of the radiation can be
inferred.
The improvements of SDDs are on the one hand due to their larger depletion
depth, which leads to a higher detection efficiency of possible X-rays from PEP-violating
transitions. On the other hand, the improved time resolution enables
the use of an active shielding against external radiation. For this purpose, 32
scintillator bars read out by 2 silicon photomultipliers each are installed around
the SDDs. Their veto signal helps to reduce the background.

\begin{figure}
\centering
\includegraphics[width=0.7\linewidth]{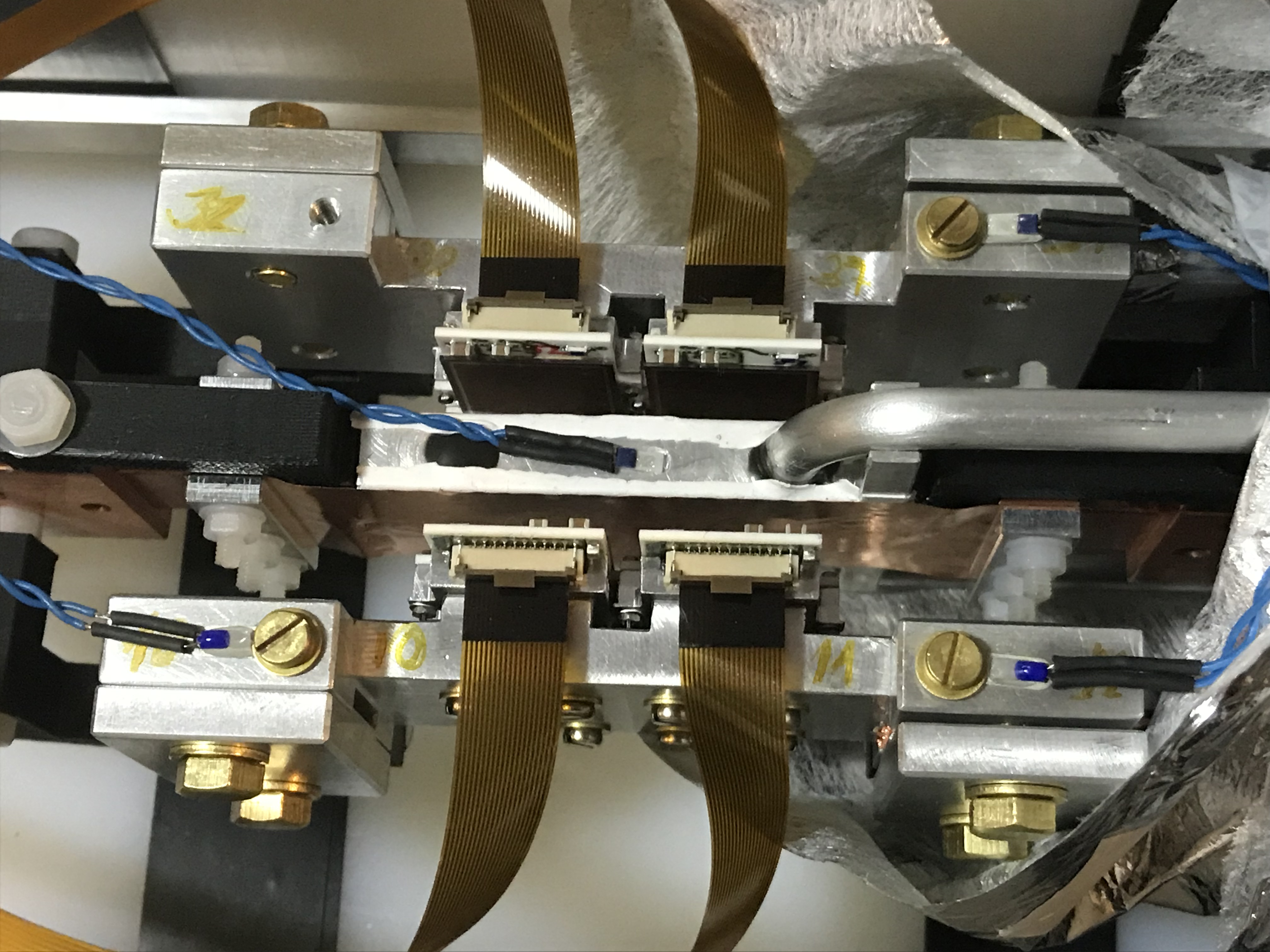}
\caption{The new Silicon Drift Detectors (4x8 SDD cells) mounted close to
the Cu target in the setup, with readout cables and
water cooling to counteract the heating due to the
high current.}
\label{fig:SDDs}
\end{figure}

\section{X-ray spectrum measured at LNGS}

The VIP2 experiment has taken over 180 days of data in the Gran Sasso underground
laboratory (LNGS). An analysis of a smaller dataset is presented in \cite{Shi2018}. With the complete dataset, the Pauli Exclusion Principle for electrons could be tested
with unprecedented precision.
\vspace*{-30mm}
\begin{figure}[ht]
\centering
\includegraphics[width=0.8\linewidth]{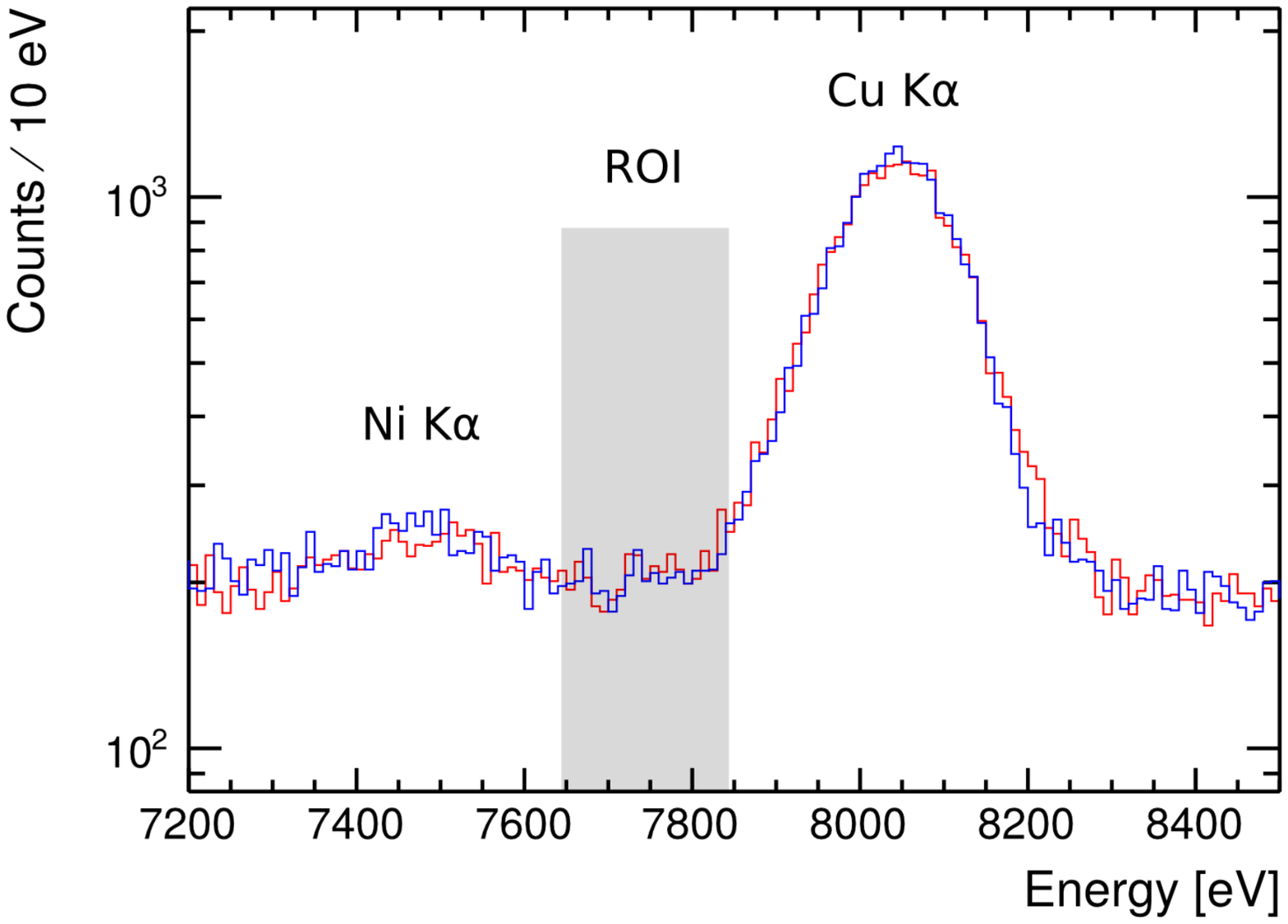}
\vspace*{-30mm}
\caption{Comparison of data taken with and without current at LNGS.
The region of interest where the PEP violating transition is
expected is marked in gray. The data corresponds to around
163 days.}
\label{fig:X-rayspectrum}
\end{figure}

 With this data taken at LNGS, a new upper
limit for the probability for a violation of
the PEP can be calculated \cite{Pichler2018}:

\begin{equation}\label{Prel-result}
  \frac{\beta^{2}}{2} \leq 1.87 \times 10^{-29}
\end{equation}

\section{Summary and outlook}
Already with the present VIP2 setup the up-to-now most stringent test of PEP 
was achieved. Presently we run the experiment with new types of SDDs providing a 
larger active area and excellent energy resolution. We will install anctive shielding based on
plastic scintillation bars coupled to silicon photomultipliers. Additionaly a optimized shielding 
will be used to further suppress background.

%\vspace*{-30mm}
\begin{figure}[ht]
\centering
\includegraphics[width=0.8\linewidth]{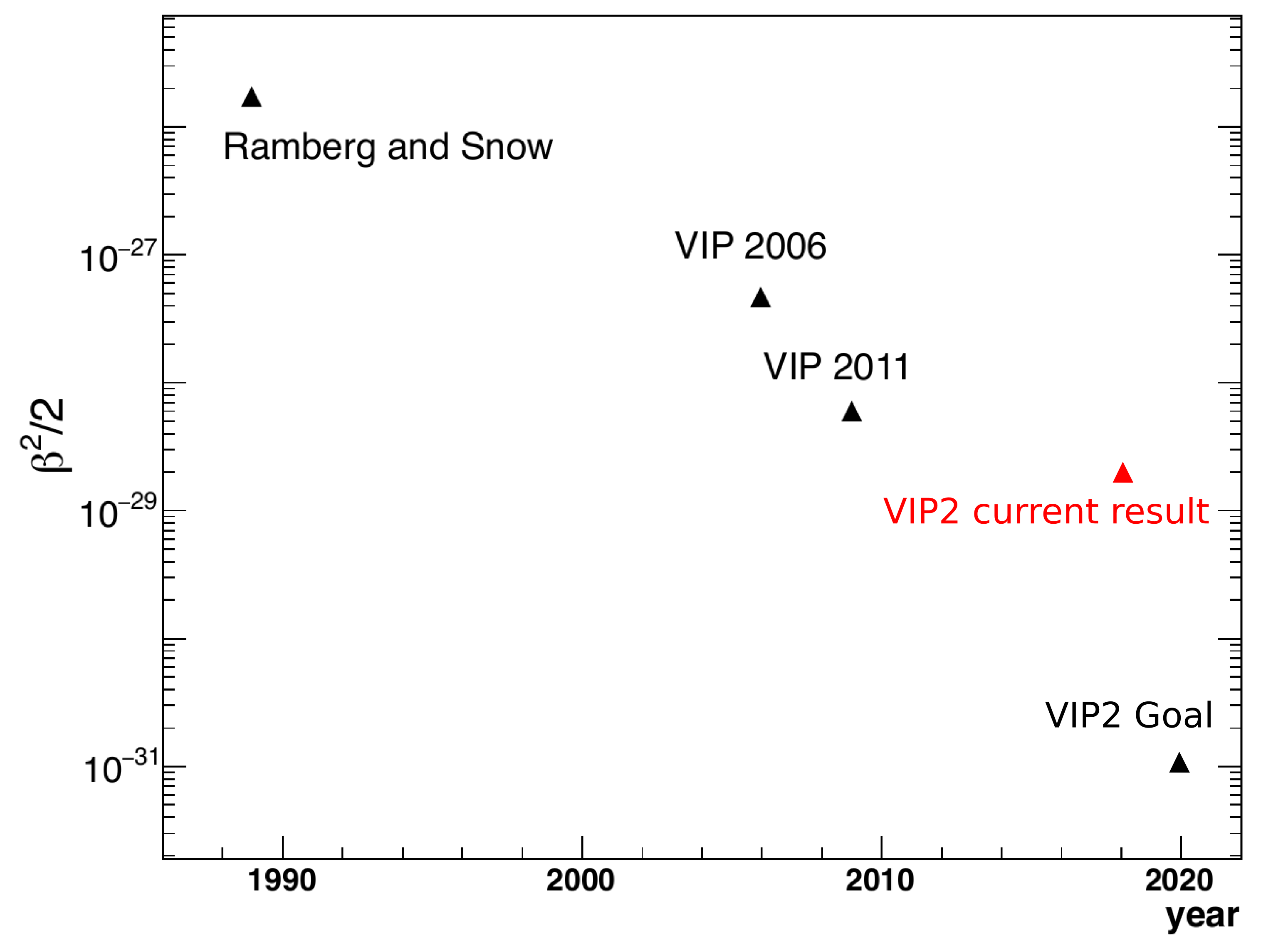}
\caption{The results of the VIP experiment, compared to the limit
we can set with the data taken by the VIP2 experiment. The
envisioned final result after the full data taking time is
shown including a planned upgrade \cite{Curceanu2017}.}
\label{fig:X-eayspectrum}
\end{figure}

With planned improvements the goal of
setting a new upper limit for the violation
of the PEP to $\sim$10$^{-31}$ could be reached after 3
years of data taking.

\section*{Acknowledgments}

We acknowledge the
very important assistance of the INFN-LNGS laboratory. We thank the Austrian Science Foundation (FWF) which supports the VIP2
project with project P 30635-N36 and W 1252-N27 (doctoral college particles and interactions)
and Centro Fermi for the grant ``Problemi aperti nella meccania quantistica''. Furthermore, these
studies were  made possible through the support of a grant from the Foundational Questions
Institute (FOXi) and a grant from the John Templeton Foundation (ID 58158).
The opinions expressed in this publication are those of the authors and do not necessarily reflect
the views of the John Templeton Foundation.

\section*{References}
%% bibliography

\end{document}